\documentclass[aip,rsi,amssymb]{revtex4-1}

\usepackage{graphicx}
\DeclareGraphicsExtensions{.eps,.pdf,.png,.jpg,.tif}
\begin{document}


\title{Optical performance of an ultra-sensitive horn-coupled transition-edge-sensor bolometer with hemispherical backshort in the far infrared}


\author{Michael D. Audley}
\author{Gert de Lange}\affiliation{SRON Netherlands Institute for Space Research, Postbus 800, 9700 AV Groningen, The Netherlands}
\author{Jian-Rong Gao}\altaffiliation{also with Kavli Institute of Nanoscience, Delft University of Technology, 
Lorentzweg 1, 2628 CJ Delft, The Netherlands}
\author{Pourya Khosropanah}
\author{Richard Hijmering}
\author{Marcel Ridder}\affiliation{SRON Netherlands Institute for Space Research, Sorbonnelaan 2, 3584 CA Utrecht, The Netherlands}
\author{Philip D. Mauskopf}\altaffiliation{current address: Arizona State University, Department of Physics, P.O. Box 871504, Tempe, AZ 85287-1504, USA}
\author{Dmitry Morozov}\affiliation{School of Physics and Astronomy, Cardiff University, Queen's Buildings, The Parade, Cardiff CF24 3AA, UK}
\author{Neil A. Trappe}
\author{Stephen Doherty}\affiliation{Department of Experimental Physics, Maynooth University, Maynooth, Co. Kildare, 
W23 F2H6, 
Ireland}

\date{\today}

\begin{abstract}
The next generation of far infrared space observatories will require extremely sensitive detectors that can be realized only by combining extremely low intrinsic noise with high optical efficiency.  We have measured the broad-band optical response of ultra-sensitive TES bolometers (NEP$\approx2\rm\ aW/\sqrt Hz$) in the 30--60-$\mu\rm m$ band where radiation is coupled to the detectors with a few-moded conical feedhorn and a hemispherical backshort.  We show that these detectors have an optical efficiency of 60\%\ (the ratio of the power detected by the TES bolometer to the total power propagating through the feedhorn).  We find that the measured optical efficiency can be understood in terms of the modes propagating through the feedhorn with the aid of a spatial mode-filtering technique.

\end{abstract}


\maketitle

\section{\label{Introduction}Introduction}

Thermal detectors that operate by measuring the heating of a sensing element due to absorbed radiation form the basis of the most versatile and sensitive detectors available to astronomers.  When combined with an appropriate radiation-coupling scheme, they can be designed to detect light from the millimeter-wave band\cite{Rich94} through X-rays\cite{XRS}.  For infrared radiation and radiation of longer wavelength, thermal detectors are used to measure the intensity of incident radiation.  In this mode they are known as bolometers.  At shorter wavelengths they are used to detect individual photons and measure their energies and in this mode they are known as calorimeters.

 A bolometer consists of an absorber, a thermometer, and a weak thermal link, of thermal conductance $G$, to a thermal bath of temperature $T_{\mathrm{bath}}$.  The absorption of radiation causes the temperature of the absorber to rise and then decay with a time constant $\tau_{\mathrm{th}}=C/G$, where $C$ is the heat capacity of the bolometer.  
 
 The transition edge sensor (TES) is a widely-used temperature sensor for bolometers\cite{Irwin2005}.  The TES consists of a superconducting film across which a constant voltage is applied.  This voltage bias results in a current through the film, which heats it so that it is in the transition region between the normal and superconducting states.  On the transition, the resistance of the film increases with temperature so that, if the TES is voltage biased, we have negative electrothermal feedback between the Joule power dissipated in the film and the temperature.  
To see how this negative electrothermal feedback occurs, suppose that the temperature of the film increases.  Then its resistance will increase. 
 Because the film is voltage biased, this increase in resistance leads to a decrease in current, and hence the Joule heating, causing the temperature of the film to decrease.  Conversely, a drop in the film resistance will cause the Joule heating to increase, warming up the film.  This means that the bias point under voltage bias is stable against temperature perturbations.  The negative electrothermal feedback also linearizes the response, supresses the Johnson noise, and shortens the time constant of the device, so that the effective time constant of the detector $\tau_{\mathrm{eff}}$ is shorter than the thermal time constant $\tau_{\mathrm{th}}$.  The sensitivity of a bolometer is characterized by the noise equivalent power (NEP), which is the power loading that would result in a signal to noise ratio of 1.  The NEP measured under dark conditions, i.e. no incident optical power, is called the dark NEP ($\mathrm{NEP_{dark}}$).  To maximize the detector's sensitivity, the NEP must be as low as possible.  The ultimate limit is the thermal fluctuation noise (or phonon noise) in the bolometer, $\mathrm{NEP_{phonon}}=\sqrt{4\gamma k T_\mathrm{c}^2 G}$, where $\gamma$ is a factor between 0.5 and 1 which accounts for the thermal 
 gradient along the link to the thermal bath.  However, in practice, there will be a trade-off between minimizing the NEP and optimizing other parameters such as the power handling and $\tau_{\mathrm{eff}}$ for the application. In order to detect incident light with high sensitivity, incoming radiation must be coupled to the detector with high efficiency.  Taking the optical efficiency $\eta$ into account, the sensitivity of the detector to incident radiation is $\mathrm{NEP_{optical}}=\mathrm{NEP_{dark}}/\eta$ in the limit of zero absorbed optical power.  
 For finite optical power, the photon shot noise of the incoming radiation will be added to the intrinsic noise of the detector.  For maximum sensitivity the intrinsic detector noise should be small compared with the photon shot noise of the background radiation.
 In this paper we demonstrate a bolometer with an ultra-low dark NEP combined with an efficient optical coupling scheme that form a highly sensitive detector of light in the 30--60-$\mu\rm m$ wavelength band. The high optical efficiency is crucial for realizing a sensitive optical detector.
 
The detectors described in this paper were developed as prototypes for the SAFARI instrument on the proposed space observatory, SPICA, which will use a large (2.5-m diameter) primary mirror cooled to $\leqslant8\rm\ K$ to observe the cold dusty Universe in the mid- and far-infrared\cite{Swinyard2009,Roelfsema2014}. The SAFARI\cite{Jackson2012} instrument is a far-infrared spectrometer that will allow us to study the dynamics and chemistry of a wide range of objects, including galaxies at redshifts up to $z=3$ and beyond. SAFARI has three detector arrays covering the wavelength ranges 
$34\mbox{--}60\rm\ \mu m$ (short wave), $60\mbox{--}110\rm\ \mu m$ (medium wave), and $110\mbox{--}210\rm\ \mu m$ (long wave), with a total of $\sim3200$ bolometers. The detectors are Transition Edge Sensor (TES) bolometers \cite{Irwin2005} using Ti/Au superconducting bilayers on thin, thermally isolated silicon nitride islands. Incoming radiation is absorbed by a 7-nm thick Ta film.
 To take advantage of SPICA's low-background cold mirror, and to meet the sensitivity requirements, SAFARI's detectors require $\mathrm{NEP_{dark}}\leqslant2\times10^{-19}\rm\ W/\sqrt Hz$, combined with an optical efficiency greater than 50\%.  To attain a phonon noise low enough to meet the requirement on $\mathrm{NEP_{dark}}$, SAFARI's TES detectors have a transition temperature, $T_{\mathrm{c}}$, of about 100~mK and are operated with a bath temperature of $T_{\mathrm{bath}}\approx50\rm\ mK$.  For high optical efficiency, the detectors use feedhorns and back-reflectors to couple light to the absorbing film.

SAFARI's detectors are over two orders of magnitude more sensitive than typical TES bolometers previously developed for ground-based applications\cite{audley2010performance} and they have correspondingly lower saturation powers (a few fW). Testing such sensitive detectors is challenging and requires careful attention to magnetic and RF shielding, stray-light exclusion, and vibration isolation. Both the TES and its SQUID readout\cite{Drung2007} are extremely sensitive to magnetic fields. Stray light exclusion is particularly important because TES detectors will saturate and become insensitive if the incident power is high enough to drive the superconducting film of the TES into the normal state.
We have built a test facility which will be used to qualify and characterize the SAFARI focal plane units and readout before they are integrated into the instrument. In addition to the strict requirements on background and interference, we require that this facility be flexible and re-configurable so that we can use it for dark and optical tests of single pixels through to the full focal-plane arrays. Through a systematic program of incremental modifications we improved the performance of the SAFARI Detector Test Facility to the point where it is now being used for routine measurements of prototype SAFARI detectors\cite{Audley2014}. 

In this paper we describe the prototype detectors and their coupling scheme and give an account of how we optimized the test facility for optical measurements.  
We have used this test facility to characterize detectors with $\mathrm{NEP_{dark}}$ as low as $6\times10^{-19}\rm\ W/\sqrt Hz$ and saturation powers below 10~fW, close to the requirements of the SAFARI flight detectors.
We present our measurements of the optical efficiency of prototype detectors with $\mathrm{NEP_{dark}}=2\times10^{-18}\rm\ W/\sqrt Hz$ in the 30--60-$\rm\mu m$ band and show how we used a spatial mode-filtering technique to allow us to interpret our results. 
Far infrared bolometers have been characterized in this wavelength range before\cite{Reveret2010,Billot2008}, although with sensitivities two orders of magnitude lower than the detectors described in this paper.
  Preliminary measurements of these SAFARI prototype detectors were performed in a test facility at Cardiff University\cite{Morozov2009}.  Prototype detectors for SAFARI's long-wave band (110--210~$\mu\rm m$) with $\mathrm{NEP_{dark}}\sim7\times10^{-19}\rm\ W/\sqrt Hz$ have been measured optically in Cambridge, UK\cite{Goldie2012}.  Other groups have reported optical measurements of far-infrared detectors in this sensitivity regime based on different detector technologies, but at longer wavelengths.  For example, optical measurements at 650~GHz of ultra-sensitive hot-electron bolometers with $\mathrm{NEP_{optical}}=3\times10^{-19}\rm\ W/\sqrt Hz$\cite{Karasik2011,Karasik2011a}, and measurements at 1.54~THz of kinetic inductance detectors with $\mathrm{NEP_{optical}}=3.8\times10^{-19}\rm\ W/\sqrt Hz$\cite{deVisser2014} have been reported.  

\section{TES Detectors and Experimental Setup}

\subsection{\label{Detectors}Detectors}
\begin{figure}
   \includegraphics[]{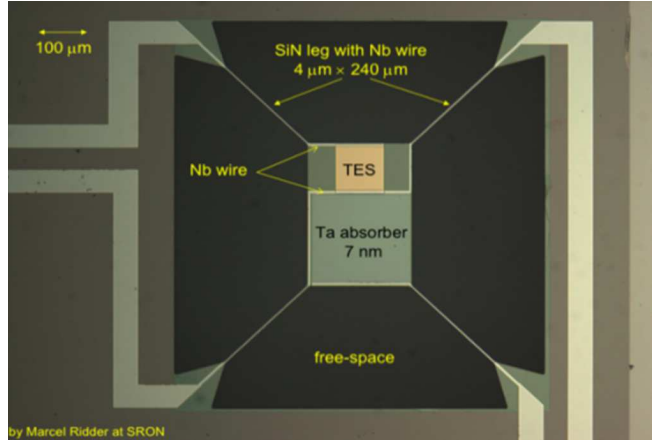}   \caption
   { \label{fig:detector}(color online) 	  
A TES bolometer of the type measured. The TES and Ta absorber sit on a $1\mbox{-}\rm\mu m$ thick SiN membrane.
The membrane is supported by four thin legs which provide a weak thermal link to the substrate and carry superconducting Nb tracks for electrical connection.  
Reproduced from Audley et~al., J. Low Temp. Physics, 176, 755 (2014). Copyright 2014, Springer.
} 
\end{figure}

We measured detectors of the type shown in Figure~\ref{fig:detector}, namely a TES bolometer fabricated from silicon nitride (SiN) membrane, with a central island and four thermally isolating straight legs.  The membrane thickness is $1\rm\ \mu m$, the width of the legs is $4\ \rm\mu m$, and the length of the legs is $240\rm\ \mu m$.  A $100\mbox{-}\rm\mu m\times100\mbox{-}\rm\mu m$ Ti/Au TES with Nb superconducting contacts is defined on the SiN island.  On the bottom part of the island is a $200\mbox{-}\rm\mu m\times200\mbox{-}\rm\mu m$ absorber made of 7-nm thick tantalum film.  The tantalum film is superconducting at the operating temperature of the TES so that it has a lower heat capacity than a normal metal film.  Lower heat capacity means a faster detector.  The radiation we are trying to detect is above the pair-breaking energy of the film and the film is engineered to have a normal-state sheet resistance close to $377\ \Omega/\square$, matching the impedance of free space.  We verified this sheet resistance by direct measurement\cite{Audley2012}.  A superconducting absorber has been used with a semiconductor temperature sensor for far-infrared bolometers before\cite{Reveret2010}, but this is the first application of a pure tantalum film as absorber.  The detectors described in this paper have moderate sensitivity (dark $\mathrm{NEP} = 1\mbox{--}2\rm\ aW/\sqrt Hz$) compared to SAFARI's requirement and, although they have been far surpassed in sensitivity by later TES
bolometers\cite{Suzuki2015,Transition-edge2013}, they absorb radiation in the same way, and are thus representative for optical characterization.

\begin{figure}
\includegraphics[]{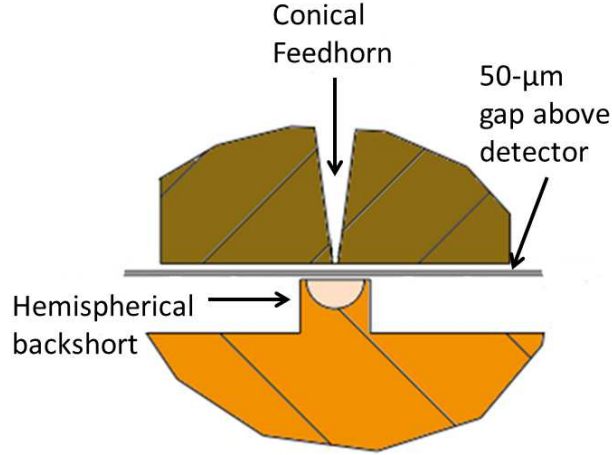}
   \caption{\label{fig:crosssection}(color online) A cross-section showing the optical coupling scheme
   \cite{Audley2014}.
   The TES bolometer sits between a conical feedhorn and a hemispherical backshort.  The exit aperture of the feedhorn is $46\rm\ \mu m$ in diameter and the diameter of the hemispherical backshort is $500\rm\ \mu m$.  Both the feedhorn and the backshort are made of copper. Reproduced from Audley et~al., Proc. SPIE 9153, 91530E (2014). Copyright 2014, Society of Photo Optical Instrumentation Engineers.
} 
\end{figure}

We tested detector chips, each of which comprises an array of five pixels.  The detector chips are mounted on a copper detector block.  The detector block has five pillars, one under each detector.  Each of these pillars has a hemispherical depression in its top that acts as a backshort.  A plate containing a single conical feedhorn is mounted on the detector block in front of the detector chip.  This horn plate can be positioned so that the feedhorn sits above any of the five detectors on a chip. For convenience we call this detector behind the feedhorn the optical detector.  The other four detectors are not illuminated and are called dark detectors.  Figure~\ref{fig:crosssection} shows the optical coupling scheme in cross-section.  The optical detector sits in front of a hemispherical backshort and behind a conical feedhorn.  The horn is 4.5~mm long and the diameters of its entrance and exit apertures are 450 and $46\rm\ \mu m$, respectively\cite{Morozov2009}.   A pair of high-pass/low-pass filters is mounted on the entrance aperture of the horn.  This optical coupling scheme has been modelled extensively\cite{Mccarthy,Trappe2012,Doherty2011} and is expected to have an ideal-case optical efficiency of about 70\%.

The detector block has superconducting feedthroughs to a printed circuit board (PCB) on its back side.  There are shunt resistors of resistance $R_{s}=1\mbox{--}5\rm\ m\Omega$ mounted on the PCB.  Each of these shunt resistors is connected in parallel with one of the TES detectors to ensure voltage biasing. There are also two single-stage SQUIDs\cite{Drung2007} mounted on the back side of the detector block so that two detectors can be read out. The SQUIDs are read out with Magnicon XXF  room-temperature electronics\cite{Magnicon}.  The Magnicon electronics has two independent channels, allowing us to read out two SQUIDs simultaneously, and providing independent DC bias to two TES detectors, so that we can measure an optical and a dark detector simultaneously.  
The Magnicon output was amplified and filtered with low-noise preamplifiers (Stanford Research Systems SR560\cite{SRS}) and recorded using an ADLINK PXI9846 analog to digital converter\cite{ADLink} (ADC).
The detector block is mounted on a table with a thermometer and heater to allow PID control of the detector-table temperature.  

\subsection{\label{Testbed}Description of the test facility}
The SAFARI Detector Test Facility is based on a Leiden Cryogenics\cite{LeidenCryogenics} dilution refrigerator with a cooling power of $\sim200\rm\ \mu W$ at 100~mK. The base temperature of the refrigerator is well below the value required for testing the SAFARI detectors ($T_{\mathrm{c}}\approx100\rm\ mK$) and gives us plenty of headroom for experiments that place additional heat loads on the system, e.g. black-body illuminators. In the current configuration, the parasitic heat loads from the optical calibration source (described in Section~\ref{BB}) limit the mixing chamber temperature to no lower than about 20~mK and $T_{\mathrm{bath}}$ to about 30~mK. The refrigerator uses no expendable cryogens and is precooled by a Cryomech PT-415 pulse-tube cooler\cite{Cryomech} which is attached to the 50-K and 3-K stages of the cryostat. To mitigate the effects of vibrations, the pulse-tube cooler has two expansion tanks and its rotary valve motor is separated from the cryostat. The expansion tanks and valve motor are mounted on the cryostat's support tripod. The main disadvantage of using mechanical cooling is that we need to protect the detectors under test from mechanical and electrical interference from the cooler. 

For detector readout and thermometry we have installed eight woven looms with 12 twisted pairs each. Two of these looms have Cu conductors for low electrical resistance (and hence low Johnson noise) and are used for the SQUID bias. The rest have CuNi conductors to minimize thermal conductance. The looms are enclosed in stainless steel tubes and heatsunk at the 50-K, 3-K, and 20-mK temperature stages of the cryostat. We have also installed stainless-steel and superconducting coaxial cables for high-frequency measurements. RF shielding is provided by two nested Faraday cages. The outer one is formed by the Dewar main shell and contains the room-temperature readout electronics as well as a multiplexer box, which allows us to connect the room-temperature readout electronics to different SQUIDs using a patchboard. The inner Faraday cage is the 3-K shield. Wires entering the 3-K shield are low-pass filtered at several hundred MHz. The system has been designed for flexibility: a reconfigurable patch board on the 20-mK stage redistributes the signals from the looms between the two experiment boxes currently installed. The experiment boxes provide a magnetically-shielded environment with ultra-low optical background for testing SAFARI's detectors.  Both of these experiment boxes are used for optical detector testing. 

\begin{figure}
\includegraphics[]{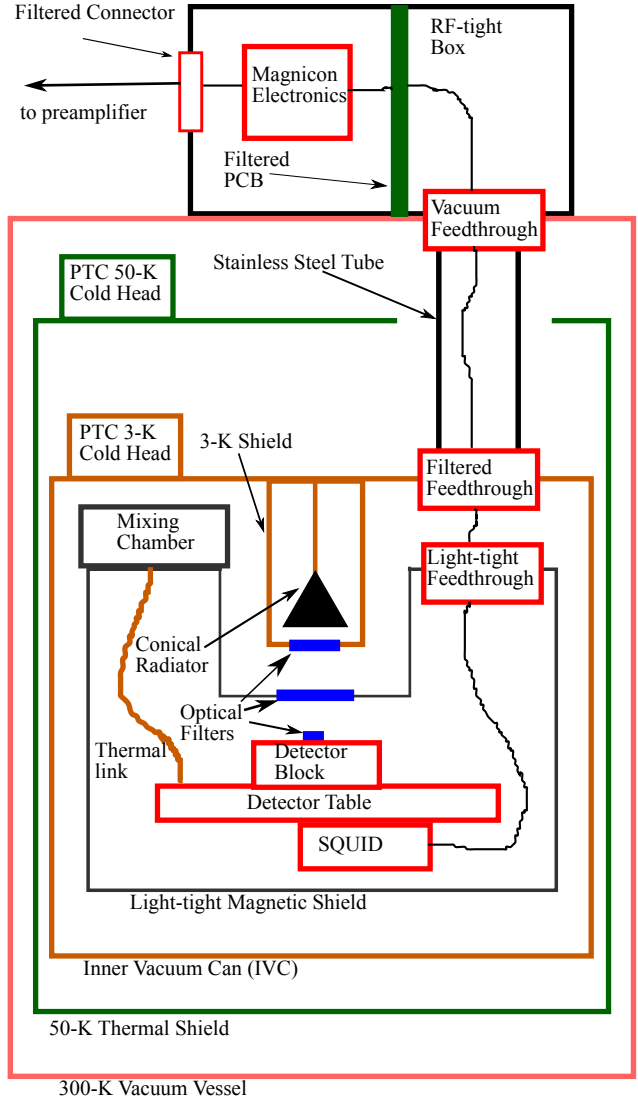}
\caption{\label{fig:testbed}(color online) Simplified schematic diagram of the test facility.  The inner vacuum can, which is heatsunk to the 3-K cold head of the pulse-tube cooler (PTC), and the main vacuum vessel form Faraday cages.  The detectors and their SQUID readout sit on a temperature-stabilized table inside a light-tight magnetic shield that is heatsunk to the mixing chamber of the dilution refrigerator. 
}
\end{figure}

Each experiment box comprises a tin-plated copper can, with light-tight feedthroughs for wiring and an absorbing labyrinth where it attaches to its base, all surrounded by a Cryoperm-10\textsuperscript{\textregistered} can\cite{Cryoperm}. We have verified that this provides good magnetic shielding and is light-tight.  The Cryoperm can shields the tin-plated can from magnetic fields as it cools through its superconducting transition so that there is no trapped flux. The copper can thermalizes more efficiently than a superconducting niobium shield would, due to its higher thermal conductivity.  
Figure~\ref{fig:testbed} shows a simplified schematic diagram of the test cryostat.  For clarity only one experiment box is shown.

\subsection{\label{BB}Black-Body Illuminator}
\begin{figure}
\includegraphics[width=0.8\textwidth]{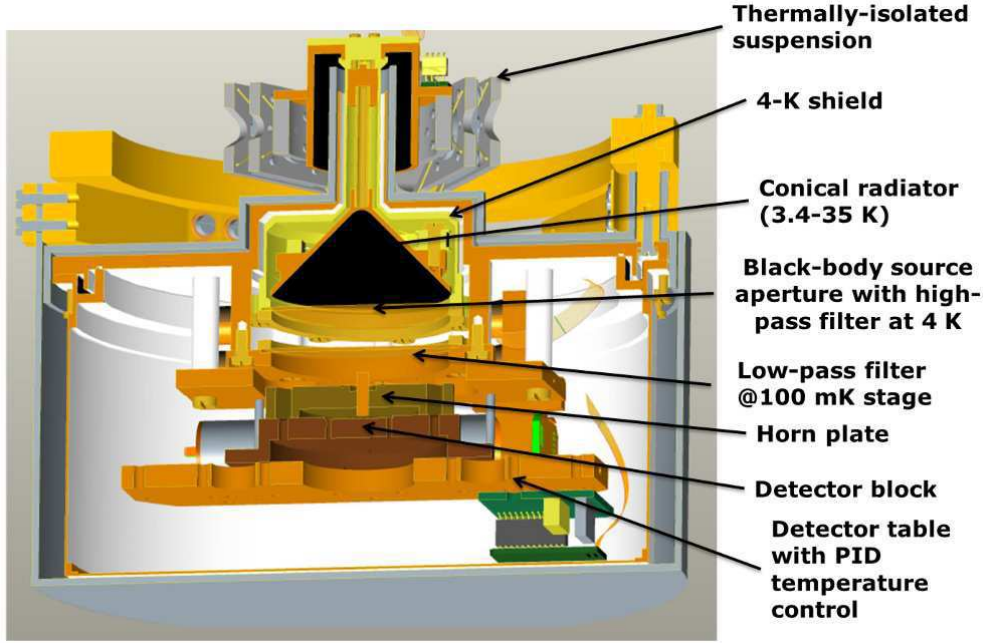}
\caption{\label{fig:BB}(color online) 
Cross-section of a light-tight, magnetically shielded experiment box.  The conical black-body radiator illuminates the detector through a 12-mm diameter aperture in the shield.  The box is heatsunk to the mixing chamber of the dilution refrigerator.
Reproduced from Audley et~al., J. Low Temp. Physics, 176, 755 (2014). Copyright 2014, Springer.
}
\end{figure}

Figure~\ref{fig:BB} shows a cross-section of an experiment box.  A conical black-body radiator is used to illuminate the detectors.  This illuminator is isolated from the 20-mK stage by a thermally isolating suspension with an intermediate stage heatsunk to the still of the dilution refrigerator.  The illuminator itself is weakly coupled to the 3-K stage.  A heater and silicon-diode thermometer are used to control its temperature in the range 3--34~K.  The upper limit on the operating temperature of the illuminator is determined by the parasitic thermal loading on the dilution refrigerator, which causes the temperature of the mixing chamber to rise.  The illuminator is surrounded by a 3-K shield with a circular aperture containing a 100-cm$^{-1}$ high-pass filter.   The light-tight superconducting can has another aperture, containing a 300-cm$^{-1}$ low-pass filter.  This aperture defines the maximum diameter of the illuminator, $D_{\mathrm{BB}}$, as seen by the detector under test.

\section{Measurements}
The basic measurement was the current-voltage characteristic curve (IV curve) that plots the TES current against the bias voltage.  From the IV curve we can derive the Joule-power vs voltage characteristic (PV curve).  When light is incident on the detector the Joule power is depressed relative to the dark PV curve.  This depression of the Joule power is equal to the absorbed optical power.  The detector was operated at an elevated bath temperature, usually $T_{\mathrm{bath}}=70\rm\ mK$, and the black-body illuminator was regulated at different temperatures in the range 3--34~K.  It was necessary to regulate the bath temperature at an elevated value because heating the black-body illuminator caused the base temperature of the refrigerator to rise.  
As the illuminator temperature $T_{\mathrm{BB}}$ increased, the temperature of the mixing chamber, $T_{\mathrm{MC}}$, rose approximately as $T_{\mathrm{BB}}^{1.7}$. $T_{\mathrm{bath}}$ was generally about 10~mK above $T_{\mathrm{MC}}$.  This meant that it was not possible to carry out optical measurements of these detectors with $T_{\mathrm{BB}}$ above about 33~K and keep $T_{\mathrm{bath}}$ below $T_{\mathrm{c}}$.

We also measured the optical photon noise in the detector and saw that it increased under optical load consistently with the absorbed optical power\cite{Audley2012}.
For noise measurements, the TES was biased at a particular point on its transition, i.e. so that the TES resistance was a particular fraction of its normal-state resistance.  The 
 current required to bias the TES at a particular point depends on the bath temperature and the loading on the detector.  We found the appropriate bias current in each case from an IV curve taken with the same bath and illuminator temperatures.  This ensured that we could measure the noise at the same TES bias point under different optical loads.  Before each noise measurement we adjusted the flux offset of the SQUID so that the output of the Magnicon XXF-1 readout was close to zero and set the gain of the preamplifier so that its output spanned a large fraction of the ADC range without saturating it.  This step ensured that we made best use of the ADC's dynamic range and minimized the effect of the ADC's intrinsic noise on the measurement.  To measure the detector noise we recorded a time series and calculated the power spectral density.  
  The preamplifier's built-in adjustable low-pass filter prevented aliasing.
 Application of the calibration derived from the IV curves gave the current noise and this was multiplied by the voltage across the TES to give the power noise.  
 This is equivalent to dividing the current noise by the zero-frequency current-to-power responsivity and the resulting power noise is equal to the NEP for frequencies that are small compared to the bandwidth of the detector.  This power noise deviates from the NEP at high frequencies because it assumes a frequency-independent responsivity, while the actual responsivity decreases with frequency for frequencies above the detector's bandwidth.  However, we can use its value at low frequencies to measure the NEP.

\begin{figure}
\includegraphics[]{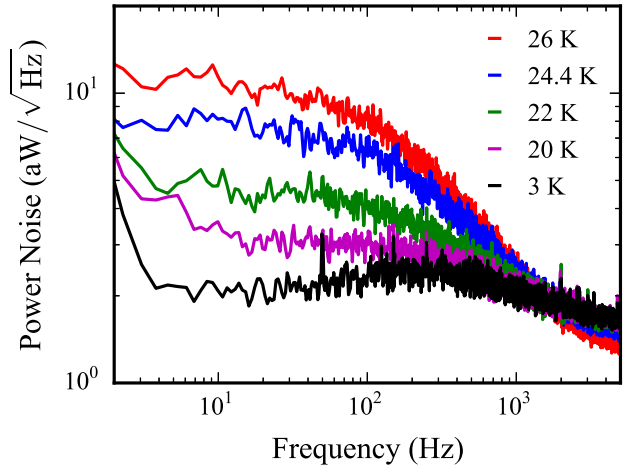}
\caption{\label{fig:PhotonNoise}(color online) Example of detector 
power noise
measured with different black-body illuminator temperatures.  
This power noise is derived by multiplying the current noise by the TES voltage, and is equal to the NEP for low frequencies (compared with the detector bandwidth).
The lowest curve with the illuminator at its base temperature 
(3~K)
shows the intrinsic noise of the detector with negligible incident optical power.  For high illuminator temperatures the photon shot noise dominates and the speed of the detector is reduced.}
\end{figure}

\begin{figure}
\includegraphics[]{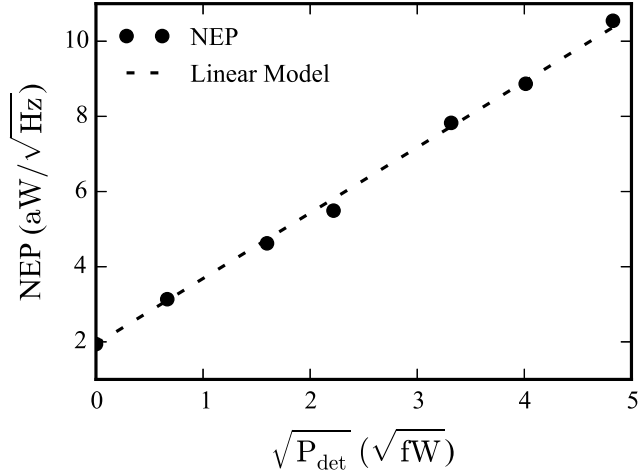}
\caption{\label{fig:NEPvsP2} Dependence of the NEP, derived from Figure~\ref{fig:PhotonNoise}, on the square root of the detected optical power, $P_{\mathrm{det}}$.  The dashed line is a linear fit showing that $\mathrm{NEP_{optical}}\propto \sqrt{P_{\mathrm{det}}}$.}
\end{figure}

An example of photon noise measuremenents is shown in Figure~\ref{fig:PhotonNoise}.  With the black-body illuminator at its base temperature, 
3~K, the intrinsic noise of the TES dominates at high frequencies. The $1/f$ noise of the SQUID and a 1.4~Hz line with harmonics from the pulse-tube-cooler dominate at low frequencies.  
When the black-body source is heated above 20~K the photon noise of the source dominates and the detector speed can be seen to decrease due to the replacement of Joule heating by optical power and the consequent reduction in electrothermal feedback.  
We calculated the NEP for each illuminator temperature by taking the mean of the power noise in the frequency range 20--24~Hz.  This was done for the data shown in Figure~\ref{fig:PhotonNoise} as well as for two illuminator temperatures, $T_{\mathrm{BB}}=23$ and 25~K, which were omitted from Figure~\ref{fig:PhotonNoise} for clarity.  Figure~\ref{fig:NEPvsP2} shows how the NEP varies with the square root of the detected optical power derived from the IV curves.  We find that the NEP is proportional to the square root of the detected power, which is consistent with photon shot noise.

While the conical horn and the hemispherical backshort are azimuthally symmetric, the detector is not, which might lead the optical efficiency to depend on polarization.  We checked the polarization dependence of the detectors by placing a polarizer between the detector and the black-body illuminator.  The absorbed optical power in the detector was measured with the polarizer in two orientations with a relative rotation of $90^\circ$.  
We found no significant difference.  
As can be seen in Figure~\ref{fig:crosssection}, there is a gap between the exit aperture of the horn and the TES detector.  We modified the horn plate to reduce this gap from 230 to $50\rm\ \mu m$ and found that the detected power increased significantly\cite{audley2013a}.  This 
is expected\cite{Mccarthy} and
can be understood intuitively in terms of a reduction of the power lost to the sides.  While the increase in optical efficiency is welcome, it highlights the sensitivity of this optical coupling scheme to the size of the gap.  This means that control of this gap is essential for ensuring the 
performance uniformity
of an array of these detectors.  
As we shall see in Section~\ref{sect:opticalefficiency} the efficiency we obtained with the 50-$\mu\rm m$ gap was high enough that we considered that the risk of making the gap too small and damaging the detector chip outweighed any further improvement in optical efficiency that might be gained from reducing the gap below 50~$\mu\rm m$ in the present setup.  Because the gap was reduced by removing material from the copper horn plate, it would also be difficult to measure the effect of intermediate gaps.
We also measured the absorbed power for similar detectors with a range of absorber sizes.  
As expected, the optical efficiency increases with absorber size.

\subsection{\label{sect:opticalefficiency}Calculation of Optical Efficiency}
\subsubsection{Single-mode Approximation}
In this work we define the optical efficiency as the ratio of power detected by the TES bolometer to the total power, $P_{inc}$,  that would propagate through an ideal version of the feedhorn.  Assuming that there are two polarizations present for each mode, the total power, $P_{\mathrm{inc}}(T_{\mathrm{BB}})$, for a black-body-source temperature of $T_{\mathrm{BB}}$ is
\begin{equation}
P_{\mathrm{inc}}(T_{\mathrm{BB}})=\int\tau(\lambda){2hc^2\over\lambda^5}{1\over e^{hc\over\lambda k T_{\mathrm{BB}}}-1}A\Omega\,d\lambda\, ,
\end{equation}
where $\tau(\lambda)$ is the transmission of the filter stack as a function of the wavelength, $\lambda$.  The area of the black-body illuminator is $A$ and it subtends a solid angle $\Omega$ to the feedhorn.
We have assumed for simplicity that only the fundamental circular-waveguide mode propagates through the horn so that the throughput per polarization is $A\Omega=A\Omega_{\mathrm{singlemode}}$ where 
\begin{equation}
A\Omega_{\mathrm{singlemode}}=\lambda^2F(\lambda,D_{\mathrm{BB}},L_{\mathrm{BB}})\, .
\end{equation}
Here $F(\lambda,D_{\mathrm{BB}},L_{\mathrm{BB}})$ is a vignetting factor that takes into account the truncation of the feedhorn's far-field beam pattern by the black-body aperture of diameter $D_{\mathrm{BB}}$ at a distance from the feedhorn of $L_{\mathrm{BB}}$.
\begin{equation}
F(\lambda,D_{\mathrm{BB}},L_{\mathrm{BB}})=1-\exp(-2(D_{\mathrm{BB}}/w(\lambda))^2)
\end{equation}
where $w(\lambda)$ is the beam radius at the illuminator aperture, based on the far-field beam pattern of the feedhorn measured at 5.3~THz, which gave us the value of $w(\lambda)$ at $\lambda=57\rm\ \mu m$.
This allows easy comparison between measurements with varying parameters.  However, as we shall see, this single-mode assumption significantly overestimates the real optical efficiency.

While a single-mode feedhorn provides the most efficient point-source coupling, an imaging bolometer array could use rectangular pyramidal-shaped horn arrays that Nyquist sample the focal plane. For this it is advantageous to use multi-moded horns that can couple all radiation in the point spread function at the focal plane to the detectors. On the other hand a multi-moded horn will have a broad beam that is susceptible to stray light and therefore needs very good cold baffling. A trade-off therefore has to be made, depending on the final architecture of the instrument.  

\subsubsection{Multi-mode Analysis}
\begin{figure}
\includegraphics[]{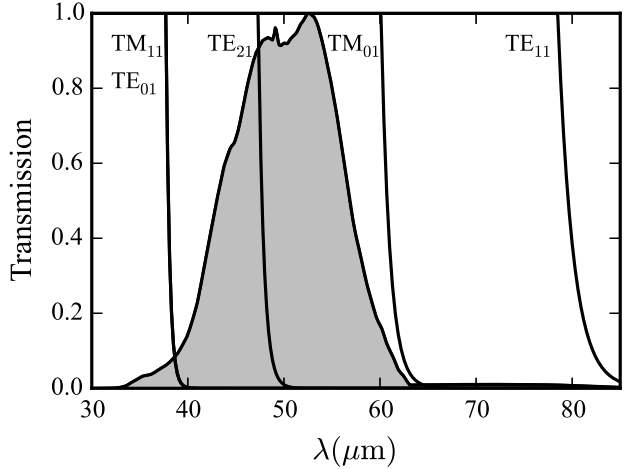}
\caption{\label{fig:modes} Transmission coefficient of the feedhorn for the first five circular waveguide modes.  Each mode can propagate in the wavelength region to the left of the corresponding transmission curve.  The shaded region represents the normalized spectrum from a 26-K black-body that has passed through the filter stack.}
\end{figure}

In order to determine which circular waveguide modes are supported by the feedhorn in the pass-band of the filter stack, we calculated the transmission of the horn as a function of wavelength for each mode.  The diameter $D(z)$ of the feedhorn varies linearly with distance $z$ from $450\rm\ \mu m$ at the mouth ($z=0$) to $46\rm\ \mu m$ at the exit aperture ($z=4.5\rm\ mm$). At a distance $z$ along the feedhorn, the logarithm of the attenuation
of radiation of wavelength $\lambda$
 produced by an infinitesimal section of waveguide of diameter $D(z)$ and length $dz$ was calculated.  The total attenuation for each mode
as a function of $\lambda$
 was 
then
found by integrating along the length of the feedhorn.  
Figure~\ref{fig:modes} shows the transmission coefficients of the first five circular-waveguide modes through the feedhorn at different wavelengths. The normalized power that would be expected from a 26-K black-body source after passing through the filter stack is also shown in order to illustrate how the regions where the circular waveguide modes are supported overlap the spectrum of the filtered black-body source.  Higher-order modes are not shown because their cut-off wavelengths are shorter than the filter stack's pass-band.  The fundamental TE$_{11}$ mode propagates across the entire pass-band of the filter stack.  The next mode, TM$_{01}$, cuts on at the long-wavelength end of the band.  However, the TM$_{01}$ mode is not expected to couple efficiently to the detector's absorbing film because it carries no power on-axis.  The TE$_{21}$ mode cuts on in the middle of the pass-band
and has two polarizations.
  The TE$_{01}$ and TM$_{11}$ modes are  supported only at the short-wave end of the pass-band, where there is very little illuminator power.  The TE$_{01}$ mode is azimuthally symmetric and thus has a single polarization.  Like the TM$_{01}$ mode, it does not carry power on axis, and is thus expected to couple poorly to the detector's 
absorbing film.  
Therefore, we can make a first-order estimate of the actual power propagating through the feedhorn and available for detection. Assuming that only the TE$_{11}$ mode, which illuminates all of the pass-band, and the TE$_{21}$ mode, which illuminates part of the pass-band, are important, we would expect to see 1--2 times as much power as in the single-mode case.
In order to estimate the optical efficiency properly, we must either have precise knowledge of  the modal content of the radiation propagating in the feedhorn, or restrict this radiation to a single mode.  The former may be calculated, although it is complicated by wavelength dependence and uncertainty in the relative power transported by the different incident modes.  
We therefore chose to take the latter approach and determine the optical efficiency with the incident reaiation restricted to a single mode.  We use the former approach, where we estimate the power carried by multiple modes,  as a consistency check.

\begin{figure}
\includegraphics[]{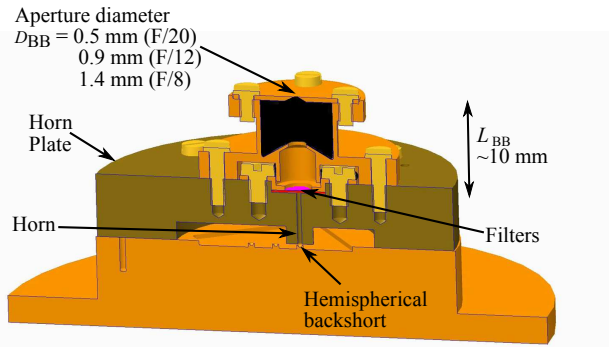}
\caption{\label{fig:can}(color online) 
Cross-section of detector holder with can for varying pinhole diameter mounted in front of the feedhorn.  The top plate contains a beam-defining pinhole and can be exchanged to change the pinhole diameter, $D_{\mathrm{BB}}$.  This allows us to make measurements with beams of different angular size to investigate the effects of higher-order modes.
}
\end{figure}

\begin{figure}
\includegraphics[]{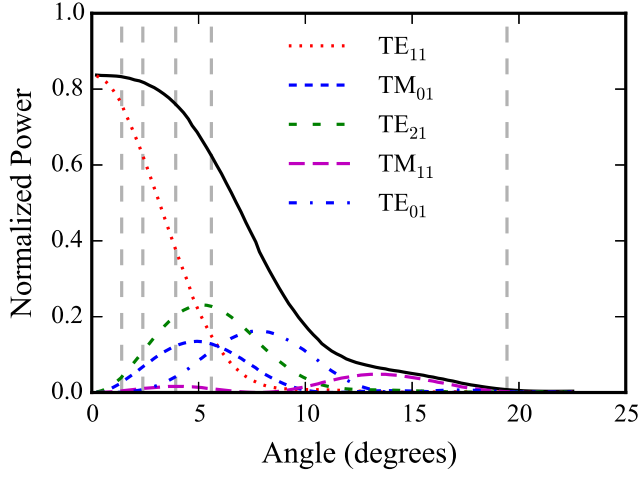}
\caption{\label{fig:beampatterns}
(color online) 
Far-field power patterns of the circular waveguide modes supported by the feedhorn in the pass-band of the filter stack.  The vertical
dashed lines show the angles subtended by the different apertures placed in front of the black-body source.  From left to right, these are 0.5, 0.9, 1.4, and 2.0, and 12~mm; the rightmost vertical 
dashed line corresponds to the case where the 12-mm diameter illuminator is unobscured and it essentially fills the beam of the horn.  Note that only the fundamental 
TE$_{11}$ mode has significant on-axis power. 
}
\end{figure}

\begin{figure}
\includegraphics[]{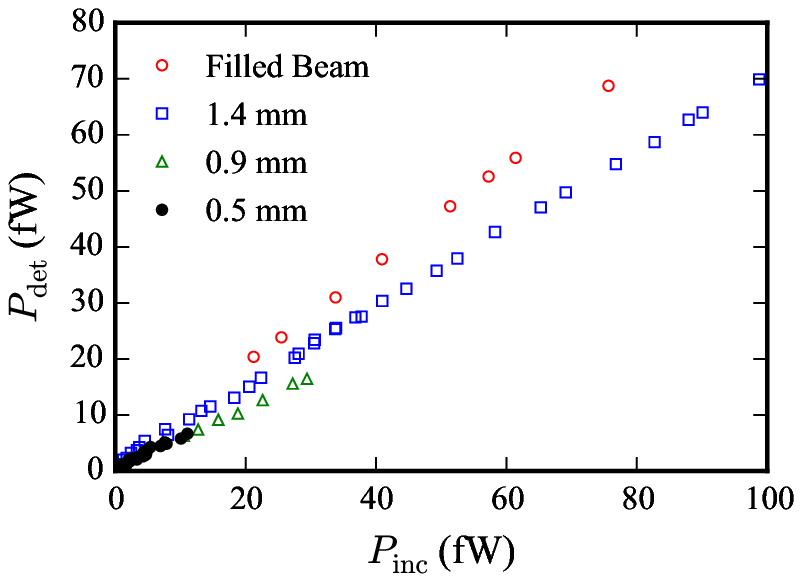}
\caption{\label{fig:pinholemeasurements}
(color online) 
Optical power detected by the bolometer, $P_{det}$, vs the incident optical power, $P_{inc}$.  Pinholes of diameter 0.5, 0.9, and 1.4~mm were placed in front of the feedhorn to reduce the angular size of the source.  The case where the source fills the beam of the feedhorn (no pinhole) is also shown.  If we assume that a single mode propagates in the feedhorn, the apparent optical efficiency is 88\% for the filled beam and 60\% for the two smallest pinholes.
}
\end{figure}

\subsection{Measurements with Spatial Mode Filtering}
We mounted a cylindrical can in front of the entrance aperture of the feedhorn, as shown in Figure~\ref{fig:can}.  We installed plates on the top of this can that had pinholes with different diameters, allowing us to measure the encircled energy of the beam of the horn with different circular apertures, or, equivalently, to make measurements with beams of different focal ratio.  Figure~\ref{fig:beampatterns} shows the calculated far-field beam patterns of the first few circular-waveguide modes\cite{Murphy1991}, with the angular sizes of the different pinholes marked.  Since the higher-order modes have mostly off-axis power, using small pinholes filters them out, so that the encircled energy of the feedhorn is then essentially the encircled energy of a single-moded horn.  Note that this holds only if few modes are present: with a higher mode content, modes of type TE$_{1n}$ and TM$_{1n}$ might be present that also carry on-axis energy.  However, of these only TM$_{11}$ is supported in the pass-band of the filter stack, and then only at the short-wave end of the pass-band.   Apertures of intermediate diameter should admit different proportions of higher-order modes according to the angles they subtend to the horn.  Figure~\ref{fig:pinholemeasurements} shows the measured optical power with different pinholes.  In order to compare the different pinholes we fit a linear model to the data and derive an effective optical efficiency under the assumption of a single mode propagating in the feedhorn.  As expected, this effective optical efficiency is erroneously high ($\sim88\%$) for the large aperture and decreases to an approximately constant value ($\sim60\%$) for the small pinholes.  In a purely single-moded horn the calculated optical efficiency should be independent of the pinhole size.  If we assume that only a single mode propagates, we find that the apparent optical efficiency is a factor of about 1.5 higher for the large aperture than for the smallest pinhole.  The estimate of 60\%\ for the optical efficiency is more consistent with the predicted ideal-case efficiency of 70\%\cite{Trappe2012,Doherty2011}.  

\begin{figure}
\includegraphics[]{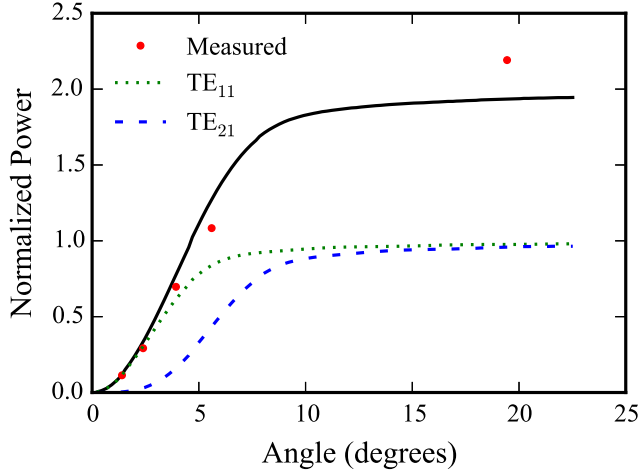}
   \caption{ \label{fig:encircledpower}(color online) 	  
Measured optical power at $T_{\mathrm{BB}}=26\rm\ K$ vs source angular size, compared with the expected encircled power derived from the far-field beam patterns of the TE$_{11}$ and TE$_{21}$ modes shown in Figure~\ref{fig:beampatterns}.  The solid line shows the sum of the encircled powers expected for the TE$_{11}$ and TE$_{21}$ modes. There is good agreement for small angular size and the model underestimates the power for the large aperture due to the off-axis contribution of higher-order modes. 
} 
\end{figure}

We calculated the encircled power seen by the detector for each pinhole diameter, when illuminated by the black-body source at a temperature of 26~K.
Figure~\ref{fig:encircledpower} shows how the measured optical power, with the black-body source at 26~K, varies with the angle subtended by the pinhole.  This is compared with the estimated encircled power from each circular-waveguide mode and with the sum of their contributions.   As expected, we see that the contribution from the fundamental 
TE$_{11}$
mode dominates for small angles, with the higher-order modes contributing at larger angles.  


%

\section{Conclusions}
We have demonstrated that an absorber-coupled, ultra-sensitive TES bolometer ($\mathrm{NEP_{dark}}\approx2\rm\ aW/\sqrt Hz$) in combination with a few-moded conical feedhorn and hemispherical backshort can detect light with high efficiency in the 30--60-$\rm\mu m$ wavelength range.
Measurements with different diameter pinholes to filter the modes 
spatially
 resulted in a consistent characterisation of the optical behavior of the detector and the contribution of higher-order modes to the optical efficiency.  This shows that the spatial-mode filtering technique is useful for analyzing the 
behavior of multi-mode detector systems.  Our measurements show that the prototype detectors for SAFARI's short-wave band have a high optical efficiency, $\eta\approx60\%$.  This high optical efficiency combined with a low dark NEP will enable SAFARI's detectors to meet their sensitivity requirements.  Although detectors combining ultra-low NEP combined with high optical efficiency in the submillimeter wavelength band have been demonstrated before\cite{Karasik2011}, our work presents the first demonstration at such short wavelengths.  Our result represents also the first optical measurement using the optical coupling configuration based on a feedhorn, an absorber, and a backshort in the short-wavelength region of the far-infrared band.  We have gained a good understanding of the optical coupling of these prototype detectors, which allows us to further optimize the coupling scheme.  For example, a flat-backshort architecture may offer a higher optical efficiency\cite{Doherty2012}, while having other advantages such as a flatter frequency response, less sensitivity to the gap between the horn and detector, and easier fabrication\cite{Withington2013}.



\begin{acknowledgments}
We thank J\"orn Beyer for providing the SQUIDs used to read out the detectors,
Douglas Griffin for providing the feedhorn, and Carole Tucker for the optical filters.  We are grateful to Stafford Withington and Peter Ade for useful discussions,
and to Darren Hayton for measuring the beam pattern of the feedhorn.
We thank Axel Detrain, Wim Horinga, Geert Keizer, Duc van Nguyen, and Willem-Jan Vreeling for their help with the test facility.  We are also grateful to the anonymous referees whose suggestions improved this paper.
\end{acknowledgments}

\bibliography{17D}

\end{document}